\documentclass[a4paper, 10pt, conference]{IEEEtran}      

\IEEEoverridecommandlockouts                              

\usepackage[a4paper,
    left=15.9mm, right=15.9mm,
    top=19.1mm,  bottom=25.4mm]{geometry}

\usepackage{amsmath} 
\usepackage{amssymb}  
\usepackage[hyphens]{url}
\usepackage{xurl}
\usepackage[hidelinks]{hyperref}
\usepackage{xspace}
\usepackage{inconsolata}

\usepackage{algorithmic}
\usepackage{graphicx}
\usepackage{textcomp}
\usepackage{xcolor}
\usepackage{array}
\usepackage{multirow}

\usepackage[T1]{fontenc}
\newboolean{showcomments}
\setboolean{showcomments}{true}
\ifthenelse{\boolean{showcomments}}
{ \newcommand{\mynote}[3]{
        \fbox{\bfseries\sffamily\scriptsize#1}
        {\small$\blacktriangleright$\textsf{\emph{\color{#3}{#2}}}$\blacktriangleleft$}}}
{ \newcommand{\mynote}[3]{}}

\newcommand{\txshare}{\textit{tx\_share}\xspace}
\newcommand{\avgtxfee}{\textit{avg\_tx\_fee}\xspace}
\newcommand{\totalgas}{\textit{total\_gas}\xspace}
\newcommand{\noofcontracts}{\textit{no\_of\_contracts}\xspace}

\newcommand{\gaspricemedian}{\textit{gas\_price\_median}\xspace}

\newcommand{\costscore}{\textit{CostScore}\xspace}
\newcommand{\efficiency}{\textit{Efficiency}\xspace}
\newcommand{\popularity}{\textit{Popularity}\xspace}
\newcommand{\pricestability}{\textit{PriceStability}\xspace}

\newcommand{\overallscore}{\textit{DeploymentScore}\xspace}
\newcommand{\normalisediqr}{\textit{Normalised IQR}\xspace}

\newcommand{\categ}{\emph}
\newcommand{\code}[1]{\texttt{\hyphenchar\font=`\-\relax #1}}

\newcounter{numobserv} 
\definecolor{beaublue}{rgb}{0.88, 0.93, 0.93}
\usepackage{tcolorbox}
\colorlet{shadecolor}{beaublue}
\tcbset{
	colback=beaublue,
	boxrule=0.5pt,
	boxsep=0pt,
	left=1pt,
	right=1pt,
	top=1pt,
	bottom=1pt,
	sharp corners,
	colframe=white,
}
\newcommand{\observ}[1]{
	\addtocounter{numobserv}{1}
	\begin{tcolorbox}	
		\textit{\textbf{Take-away\,\thenumobserv\,:} #1 }	
	\end{tcolorbox}
}

\title{\LARGE \bf
Towards an Intelligent Decision Engine for Smart Contract Deployment
}

\author{
\IEEEauthorblockN{Oana-Otilia Captarencu}
\IEEEauthorblockA{\textit{Faculty of Computer Science} \\
\textit{Alexandru Ioan Cuza University}\\
Ia{\c{s}}i, Romania \\
oana.captarencu@info.uaic.ro}
\and
\IEEEauthorblockN{Emanuel Onica}
\IEEEauthorblockA{\textit{Faculty of Computer Science} \\
\textit{Alexandru Ioan Cuza University}\\
Ia{\c{s}}i, Romania \\
emanuel.onica@info.uaic.ro}
\and
\IEEEauthorblockN{Andrei Arusoaie}
\IEEEauthorblockA{\textit{Faculty of Computer Science} \\
\textit{Alexandru Ioan Cuza University}\\
Ia{\c{s}}i, Romania \\
andrei.arusoaie@info.uaic.ro}
}

\begin{document}

\maketitle
\thispagestyle{empty}
\pagestyle{empty}

\begin{abstract}
Layer 2 (L2) blockchain solutions emerged in recent years, predominantly in the Ethereum ecosystem, as a viable option for scaling a main blockchain network (L1). 
However, the selection of a L2 solution for deploying smart contracts backing blockchain applications is a decision that depends on many factors, such as costs, efficiency, popularity, and stability.
We propose here a systematic methodology to help with this decision making process by using a set of relevant metrics and introducing user-defined profiles that capture different priorities among these metrics.
We collect the major categories of smart contracts from a community-maintained data source, and we gather transactions for a six months period from three major L2 solutions: Arbitrum, Optimism, and Base.
For each contract category and each blockchain, we compute a set of metrics and scores tailored to each user profile. 
By applying our methodology, we identify a clear bias in the classification, showing which L2 solutions are better suited for particular application types and user profiles.

\end{abstract}

\section{Introduction}

Blockchain networks are distributed platforms that maintain a replicated ledger of immutable transactions, stored in securely chained blocks. 
Ethereum~\cite{wood2014ethereum,buterin2014ethereum} was the first blockchain to enable the deployment of smart contracts, small programs that run on the network's nodes. 
The smart contracts can power a variety of applications that benefit from the network's decentralised consensus.
Ethereum remains the dominant ecosystem for smart contract development, largely due to its popular tooling and large developer community. 
Once deployed, a smart contract exposes functions that can be called directly and executed in a decentralised manner. 
Each call that modifies the contracts' state is, in fact, a transaction, and a fee is charged to prevent network abuse and denial of service attacks.

Ethereum is subject to well-known scalability limitations, particularly regarding transaction fees and the number of transactions it can process per second.
To address these limitations Layer~2 (L2) solutions have emerged, with~\emph{rollups} representing the dominant approach. 
Instead of replacing the Ethereum mainnet, rollups execute transactions and periodically submit resulting state updates to Ethereum's main chain (L1). 
Thus, they retain the security guarantees of L1 while significantly increasing throughput and lowering costs. 

From a developer perspective, one of the most significant advantages of L2 rollups is their compatibility with the existing Ethereum toolchain: smart contracts can be written, tested, and deployed on L2 in the same way as on L1.
However, L2 rollups differ in how they compute transaction fees. 
While L1 fees depend directly on the bytecode of executed contracts, L2 fee models must also account for the cost of posting transaction data to L1. 
Therefore, L2 rollups introduce non-standard fee variations, leading to cost fluctuations that are not always transparent to developers or even easy to understand. 
Still, L1 is considerably more volatile than rollups~\cite{DBLP:conf/debs/OnicaBACA25}, but at the same time, there are higher security risks on L2~\cite{DBLP:journals/corr/abs-2504-10535}. 

All these raise an important question: \emph{given a smart contract, which L2 network is the most suitable deployment target}? 
Smart contracts are not uniform in their behaviour: a contract implementing a token standard, 
a decentralised exchange (DEX), or a lending protocol, each exhibits a distinct pattern of function calls, storage access, and computational load. 
Therefore, finding the best deployment target requires understanding how different smart contract categories are affected by each network's costs, efficiency, popularity, and stability.
So far, there is no established methodology for finding an answer, and the lack of properly curated data makes decision-making hard.
In our work, we propose such a methodology and evaluate its capability towards providing a deployment choice among three of the most popular rollups.

\paragraph*{Contributions} The main contributions of this paper are:
\begin{enumerate}
    \item a methodology for identifying an L2 rollup to deploy a smart contract, using relevant network metrics; 
    \item establishing a set of user profiles that weight these metrics for a multi-faceted evaluation of L2 solutions;
    \item extraction and aggregation of a list of smart contract categories from a community-maintained data source;
    \item extraction of six months of transaction data from three major L2 rollups — Arbitrum, Optimism, and Base; 
    \item evaluation of our methodology by computing scores for each metric per contract category, and a comparative analysis of the results. 
\end{enumerate}


\paragraph*{Paper Organisation}
Section~\ref{sec:background} presents background information on the L2 Ethereum ecosystem and related work. Section~\ref{sec:methodology} introduces our systematic methodology, including the collection of smart contract categories and the main metrics we use. In Section~\ref{sec:results}, we identify three user profiles and, for each profile, the overall scores per category along with the key findings. Finally, we conclude in Section~\ref{sec:conclusions}.

\section{Background and Related Work} 
\label{sec:background}
In this section, we cover background concepts on smart contract costs and the Ethereum ecosystem, followed by an overview of work related to selecting a deployment medium.

\subsection{Costs and Ecosystem Overview}
Smart contract functions, corresponding to transactions, are executed in the Ethereum ecosystem by the core component of a network node, the Ethereum Virtual Machine (EVM). 
The transaction cost is quantified in units of~\emph{gas}.
This is defined by a standard specification that accounts for the computational and space complexity of the transaction submission and execution~\cite{wood2014ethereum}.
As a result, a given contract function typically incurs similar gas costs across EVM-compatible networks, although differences may arise due to protocol changes or different EVM implementations. 

However, what can vary across networks is the price set per gas unit.  
This is used to compute a transaction's fee, which is the amount effectively paid in Ether (ETH), the Ethereum cryptocurrency, typically expressed in smaller denominations, such as Gwei (1 ETH = $10^9$ Gwei).
The precise fee formula varies across networks, with its main component computed as the product of the transaction gas cost and the gas price.
The latter is typically split into a base fee burned by the network and a priority fee to incentivise faster processing.
Transactions are included in securely chained blocks, which have a gas limit and are produced at a network-dependent frequency. 
The fee levels that compose the gas price vary dynamically based on network demand and competition for~block~space.

As already mentioned, L2 solutions are part of a broader ecosystem that builds on top of L1 infrastructure.
The two main technical approaches of L2 networks are optimistic rollups and zero-knowledge (ZK) rollups. 
Optimistic rollups, such as Arbitrum~\cite{arbitrumnitro2022} and Optimism~\cite{optimismspec}, assume transactions are valid by default and rely on a fraud-proof challenge period to identify invalid state transitions submitted to L1. 
This design can support significantly higher throughput than L1 while keeping gas prices well below those on the Ethereum mainnet. 
By contrast, ZK rollups, with zkSync Era~\cite{zksyncprotocol} as a prominent example, generate a cryptographic proof attesting to the correctness of the state transition resulting from a batch of transactions when submitting to L1. 
While this enables faster transaction settlement, it incurs higher costs than the optimistic approach, which is currently reflected in the reduced popularity of these networks.

Each L2 solution comes with its own fee model, performance characteristics, and methods for mitigating fraud. 
To make informed decisions about where to deploy a smart contract, a careful analysis of existing data is required.
A widely known source for blockchain analytics is Dune Analytics~\cite{dune}, which provides access to blockchain data (e.g., fresh transaction receipts) that can be queried and visualised. 
In addition, it offers a community-maintained labelling ontology for the stored data.
However, this data alone is insufficient to support deployment decisions. 
The data is raw, and the existing labelling ontology is sometimes inconsistent. 
Therefore, identifying the relevant information corresponding to decisional factors is nontrivial.

\subsection{Related Work}
A complementary line of research~\cite{11113254} investigates optimisation strategies from an architectural perspective, focusing on L2 solutions, off-chain storage, and smart contract optimisations. 
Unlike our work, it evaluates whether to use L2 solutions rather than which one to select.

Other studies, such as~\cite{dyade2025eip4844}, focus on an in-depth analysis of the economic impact of EIP-4844, a recent standard concerning L1 data submission, examining gas cost reductions and assessing the upgrade's implications. However, they do not address the question of where to deploy a smart contract.

More closely related to our research are some empirical comparisons of L2 solutions hosted on various websites.
For example, the study in~\cite{markaicode2025} presents performance benchmarks for different L2s, including metrics such as maximum and sustained throughput (TPS), initial confirmation time, finality, and the cost of typical transactions (e.g., token swaps or NFT minting). 
While these metrics are relevant for decision-making, such studies are not scientific in nature, lack clear authorship, and are often not reproducible.

A more reliable source of data for decision-making is provided by analytics platforms such as GrowThePie~\cite{growthepie}, L2Beat~\cite{l2beat}, and Dune Analytics~\cite{dune}. These platforms offer a wide range of metrics and dashboards for analysing L2 blockchains. 
However, they primarily focus on data visualisation and exploration, while performing custom analyses may require technical expertise for data extraction, potential access costs, and additional processing effort. 
These platforms give network-level aggregates, not per-contract-category breakdowns. Also, they do not offer profile-weighted scoring that combines cost/popularity/efficiency/stability into a single deployment decision.

In this paper, we propose a methodology that systematically leverages such data sources.
In particular, we use data obtained from Dune Analytics, while developing our own framework and scripts to process this data in a meaningful and reproducible manner.

\section{Methodology}
\label{sec:methodology}

Our methodology begins by extracting transactions to smart contracts over a six-month period, from September 2025 until March 2026, across the currently most popular three optimistic rollups: Optimism, Arbitrum, and Base~\cite{l2beat}. 
We identify the smart contract categories, and subsequently we leverage Dune’s SQL query mechanism to extract relevant data for \emph{each contract category} and \emph{each network}:

\begin{itemize}
    \item \txshare: the total number of transactions;
    \item \avgtxfee: the average transaction fee;
    \item \totalgas: the total gas consumption; and
    \item \noofcontracts: the number of contracts.
\end{itemize}

Following this data extraction, we compute several key metrics derived from this information. 
We then analyse the obtained results and 
draw useful conclusions on deployment.

\subsection{Identifying Smart Contract Categories}

We initially attempted to identify relevant smart contract categories within Dune data by leveraging the \code{labels.addresses} table, which associates contract addresses with descriptive category labels. 
However, the categorisation we found in \code{labels.addresses} proved too broad, encompassing highly generic categories like \categ{contracts}, \categ{infrastructure}, \categ{institution}, and \categ{dex}.
We also identified several categories irrelevant for the situation of a contract having a deployment choice, such as \categ{rollup} (internal system operation contracts) 
or \categ{bridge} (transfers between specific networks).
We only retained \categ{CEX} (centralised exchanges) and \categ{NFT} (non-fungible tokens) as specific enough categories for custom applications using contracts. 
We considered that~the~broader categories \categ{contracts}, \categ{infrastructure}, \categ{institution}, and \categ{dex} can be further refined, and thus,~we~investigated~the \code{labels.owner\_details} and \code{labels.owner\_addresses}~data~sources.
These tables contain details on the contract owners, allowing us to infer more specific categories like \categ{Lending}, \categ{Derivatives},~\categ{Creators and Brands}, \categ{Custody Services},~and~others. 
Following this process, we reached 24 refined categories that enable a more targeted analysis of the smart contract landscape.


\begin{figure*}
    \centering
    \includegraphics[width=\textwidth]{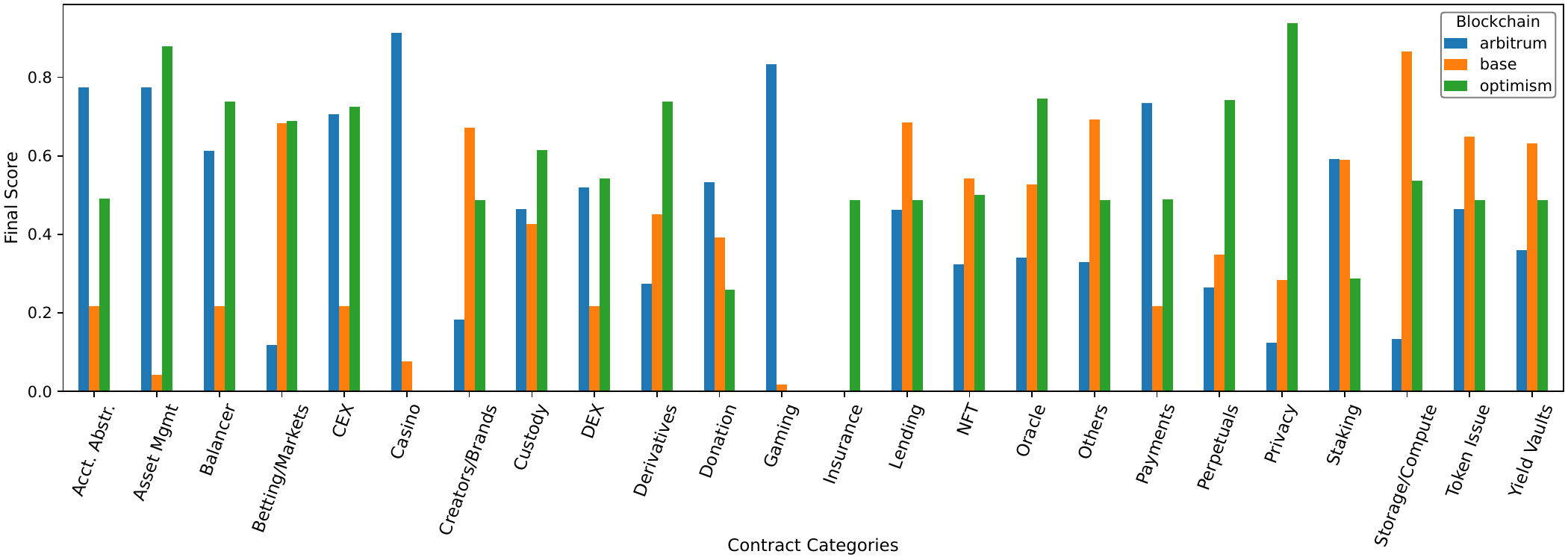}
    \caption{Final scores for the \emph{Cost-focused user} profile for each network -- Arbitrum, Base, and Optimism -- across multiple contract categories.}
    \label{fig:cost-focused}
\end{figure*}

\subsection{Relevant Metrics}
\noindent
Using the aforementioned \txshare, \avgtxfee, \totalgas, and \noofcontracts statistics extracted from the transactions corresponding to a contract category, we compute several metrics discussed below.
Deriving these metrics at the category level enables a more rigorous evaluation of the suitability of different rollups.
To enable meaningful comparisons between data that can vary significantly, we use a min-max normalisation function:
\begin{equation}
| \textit{series} | = \frac{\textit{series} - \textit{min}(\textit{series})}{\textit{max}(\textit{series}) - \textit{min}(\textit{series})},    
\end{equation}
where~\emph{series} is represented by the three rollup values.~The~min-max normalisation has an advantage that is relevant~for our approach: the ability to combine various~metrics that~are measured on different scales in each network.~This~part is~essential when we  define user profiles that combine~gas~consumption,~the number of transactions or the average transactions fees.

\subsubsection{Cost Efficiency Score}

This metric transforms the average transaction fee into a score, where a higher score indicates a lower average transaction fee. 
This is done by inverting the normalised value of the average tx fee:
\begin{equation}
\costscore = 1 - \mid\avgtxfee \mid    
\end{equation}

\subsubsection{Efficiency Score}
The efficiency score is the number of processed transactions per consumed gas. 
We compute it as total transactions over total gas usage, and then normalise it:
\begin{equation}
\efficiency = \left| {\txshare}/{\totalgas} \right|
\end{equation}

\noindent
Higher values indicate more efficient use of gas resources.

\subsubsection{Popularity Score}

The popularity score is defined by:
\begin{equation}
\popularity = w_1 \cdot |\txshare| + w_2 \cdot |\noofcontracts|,    
\end{equation}

\noindent
where \(w_1=0.3\) and \(w_2=0.7\) are the weights that control the relative importance of transaction count and the number of deployed contracts.
Normalisation is needed here because \txshare is way greater than \noofcontracts. 
Higher values of the popularity metric indicate that the rollup under question is more widely adopted for the considered category. 

\subsubsection{Stability Score}
This metric uses overall gas price variability to assess the expected stability of a particular network in terms of transaction submission costs. 
For this purpose, we use the Interquartile Range (IQR) of gas prices recorded on each network, measured over a 30-day sliding window with a 10-day step, across the monitored interval. 
We observe that gas prices follow different scales across networks.
We are interested in stability here, not in absolute value differences (which are already captured in the cost efficiency score). 
Therefore, we perform a preliminary IQR normalisation in order to obtain comparable values on fluctuation, by dividing each IQR by its network's median price:
\begin{equation}
\normalisediqr = {IQR}/{\gaspricemedian}    
\end{equation}
A lower normalised IQR indicates less fluctuation.
Finally, to scale a subunitary stability score, we apply the formula below, where a higher result implies more predictable costs.
\begin{equation}
\pricestability = {1}/{(1 + \normalisediqr)}    \end{equation}
\subsection{Deployment Decision Formula}

The final deployment decision formula that we propose is:
\begin{equation}
    \begin{aligned}
\overallscore =~& \alpha \cdot \costscore\!\! & \mathop{+}  ~& \beta \cdot \popularity  \\
\mathop{+}~&\gamma \cdot \efficiency \!\!& \mathop{+} ~& \delta \cdot \pricestability
\end{aligned}
\end{equation}

\noindent
where \(\alpha\), \(\beta\), \(\gamma\), and \(\delta\) represent the weights for \emph{cost sensitivity}, \emph{ecosystem maturity/popularity}, \emph{execution efficiency}, and  \emph{fee predictability}, respectively. 
Our deployment decision formula is an instance of Multi-Criteria Decision Making (MCDM), namely Simple Additive Weighting (SAW)~\cite{hwangyoon1981mcdm}, the most direct MCDM aggregation scheme, where alternatives are ranked against several weighted criteria.

\begin{figure*}
    \centering
    \includegraphics[width=\textwidth]{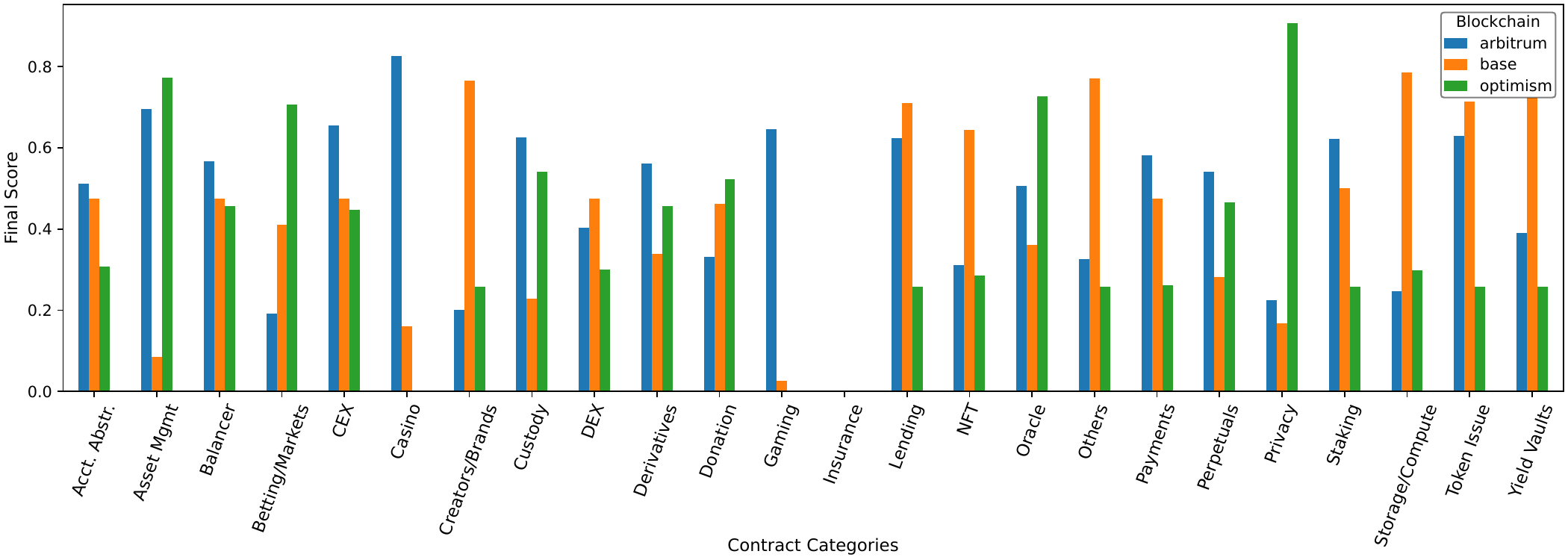}
    \caption{Final scores for the \emph{Ecosystem/maturity-focused user} profile for each network -- Arbitrum, Base, and Optimism -- across multiple contract categories.}
    \label{fig:maturity}
\end{figure*}

\section{Results}
\label{sec:results}


\subsection{User profiles}
Based on the values of \(\alpha\), \(\beta\), \(\gamma\), and \(\delta\), we can establish several user profiles, which are evaluated for each category and each rollup. The values are chosen by typical usage scenarios and relative importance observed in practice. 

\subsubsection{Cost-focused user} This profile prioritises the minimisation of deployment and execution costs, while efficiency and popularity remain relevant but secondary, and price stability has a lower influence:
\begin{equation}
(\alpha,\beta,\gamma,\delta) = (0.45, 0.2, 0.25, 0.10).
\end{equation}

\subsubsection{Ecosystem/maturity-focused user}
This profile corresponds to users who value established contract utility and broad adoption (popularity) in that category, while the other weights are almost equal to each other:
\begin{equation}
(\alpha,\beta,\gamma,\delta) = (0.2, 0.45, 0.2, 0.15).
\end{equation}

\subsubsection{Balanced/pragmatic user}
This profile reflects a general-purpose strategy without focusing on a single dominant priority, with a slight preference given to cost and popularity/adoption:
\begin{equation}
(\alpha,\beta,\gamma,\delta) = (0.3, 0.3, 0.25, 0.15).
\end{equation}

\subsection{Results for the \emph{Cost-focused user} profile}

The final results for the \emph{Cost-focused user} profile are shown in Figure~\ref{fig:cost-focused}. 
For several categories, there are clear winners in terms of cost effectiveness.
Arbitrum performs best for \categ{Account Abstraction}, \categ{Casino}, and \categ{Gaming}, while Optimism dominates the \categ{Privacy} and \categ{Insurance} categories. 
The data support Arbitrum’s claims of providing effective solutions for gaming and account abstraction through instant finality and high throughput.
Optimism leads in \categ{Privacy} and \categ{Insurance} due to its low \avgtxfee in these categories and also due to the fact that it inherits safety and decentralization from the Ethereum chain, via its fraud proof mechanisms.

We observe that Arbitrum and Optimism obtain better results than Base for \categ{Asset Management}, \emph{Balancer}, \emph{Centralized Exchange} (CEX), and \categ{Decentralized Exchange} (DEX), with Optimism having a significant lead for \categ{Privacy Services}.
On the other hand, there are categories where Base outperforms~the other two rollups, though the difference is not large. This is the case for \categ{Creators and Brands}, \categ{Lending}, \categ{NFT}, \categ{Storage/Compute}, \categ{Token Issue}, and \categ{Yield Vaults}.~Applications in these categories typically generate numerous structurally similar operations, including minting, transferring, approving, claiming, and updating token ownership. These repetitive~transaction patterns can benefit from Base’s batching and compression mechanisms, which amortize data-publication costs across multiple transactions.

\observ{
We notice that the investigated rollups are relatively well balanced in terms of dominating the categories w.r.t. costs.
This split indicates that certain categories of contracts tend to cluster on certain rollups where they gain~a cost advantage.
This correlates with observations in our maturity profile evaluation.
}

\begin{figure*}[t]
    \centering
    \includegraphics[width=\textwidth]{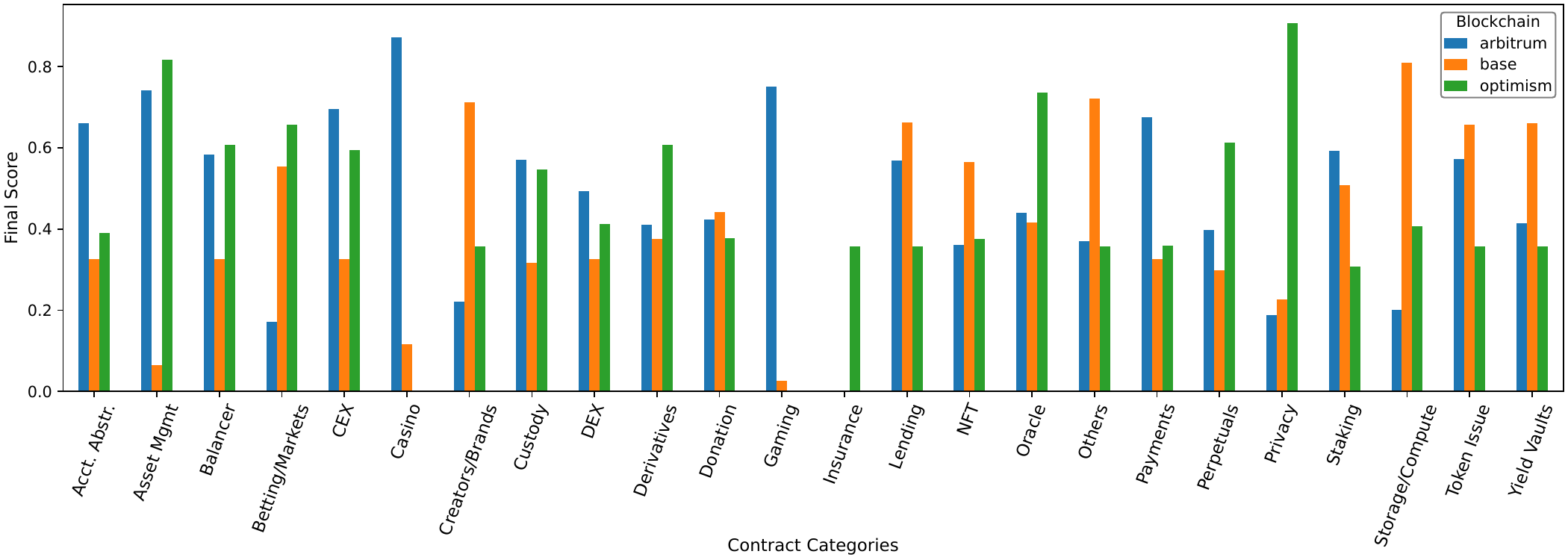}
    \caption{Final scores for the \emph{Balanced/pragmatic user} profile for each network -- Arbitrum, Base, and Optimism -- accross multiple contract categories.}
    \label{fig:balanced-user}
\end{figure*}

\subsection{Results for the \emph{Ecosystem/maturity-focused user} profile}

The scores for the \emph{ecosystem/maturity-focused user} profile are shown in Figure~\ref{fig:maturity}. 
Overall, the rankings across most categories are similar to those observed in Figure~\ref{fig:cost-focused}.
However, we observe some differences. 
In the \categ{Account Abstraction} and \categ{Payment Processing} categories, Base outperforms Optimism, although it still remains behind Arbitrum. 
Then, in the \categ{Donation} category, Optimism moves into the leading position.
Furthermore, Arbitrum takes the lead in \categ{Custody Services}, \categ{Derivatives}, and \categ{Perpetuals}, suggesting a stronger adoption. 

Overall, Arbitrum shows consistent improvement in DeFi-related categories compared to the cost scores. 
Base also exhibits a slight advantage over Optimism in more niche categories (e.g. \categ{Account Abstraction} and \categ{Payment Processing}).

\observ{
For many categories 
the rankings follow the same trend as those observed under the cost-focused scores.
This confirms an expected correlation between lower costs and higher popularity.
However, the maturity profile scores also exhibit noticeable positive differences (w.r.t. cost-oriented scores) for certain categories. 
This indicates that, even when costs are not the best, some rollups remain popular choices for particular types of smart contracts.
}

\subsection{Results for the \emph{Balanced/pragmatic user} profile}

The results for this profile are presented in Figure~\ref{fig:balanced-user}. Considering the results obtained for the other two profiles, several patterns emerge that provide additional insights.

First, for some categories, the clear network winners can be identified. 
Arbitrum appears to be the best choice for \categ{Account Abstraction}, \categ{Casino}, \categ{Gaming}, and \categ{Payment Services}. 
Optimism performs best in \categ{Insurance}, \categ{Oracle}, and \categ{Privacy Services}, while Base leads in \categ{Creators and Brands}, and \categ{Decentralized Storage and Compute}.

Second, in several categories, certain rollups remain leaders but are closely followed by their competitors. 
Arbitrum leads in \categ{Centralized Exchange}, \categ{Custody Services}, and \categ{Decentralized Exchange}, closely followed by Optimism. 
In the \categ{Staking} category, Arbitrum is followed closely by Base.

Similarly, Optimism leads in \categ{Asset Management}, \categ{Balancer}, and \categ{Perpetuals}, with Arbitrum as a close competitor. 
Optimism also leads in \categ{Sports Betting \& Prediction Markets}, followed by Base, and in \categ{Derivatives}, where it is closely followed by both Arbitrum and Base.

Finally, Base is the leading rollup in \categ{Lending}, \categ{NFT}, \categ{Token Issuance}, and \categ{Yield Vaults}, although it is closely followed by the other two platforms.

\observ{
The balanced profile typically brings two of the three rollups closer together, distributing leadership equally among categories, with 8 led by Arbitrum, 8 by Optimism, and 7 by Base.
Arbitrum appears well suited for DeFi and high-activity categories, such as gaming and payment services.
Optimism stands out in specialized services, including privacy services and insurance, and also competes with Arbitrum in centralized and decentralized exchanges as well as derivatives.
Base is well suited for creator ecosystems, contracts, and token-related applications, as well as niche use cases such as lending and yield vaults. 
}

\subsection{Gas price evolution}

\begin{figure*}[ht]
  \centering
  \includegraphics[width=\textwidth]{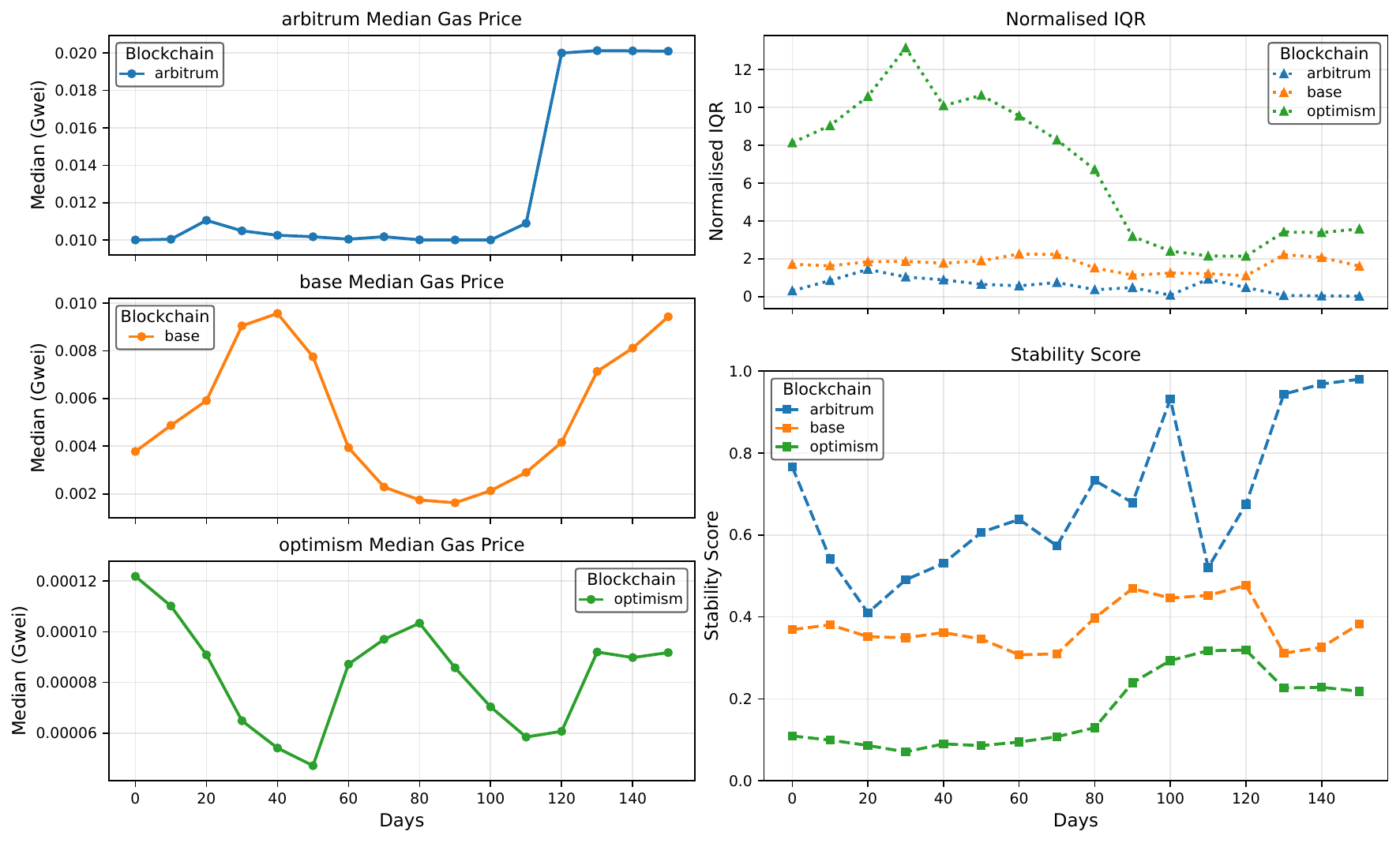}
  \caption{\label{fig:price_evolution}The evolution of gas price across the considered rollups: Arbitrum, Base and Optimism. Median gas prices expressed in Gwei are shown in the left, the normalised IQR in the top-right, and the derived stability scores in the bottom right.}
\end{figure*}

Given the importance of price in the decision-making process, independent of the user profile, we further analyse its evolution over the period under investigation.
In Figure~\ref{fig:price_evolution}, we show the time evolution of median gas prices, the normalised IQR, and the resulting stability scores for Arbitrum, Base, and Optimism. 
The results highlight differences in fee dynamics across the considered rollups.

Arbitrum maintains relatively stable median gas prices for most of the period under consideration, followed by a noticeable increase toward the end.
Base exhibits a cyclical pattern, with alternating periods of rising and falling values, while Optimism consistently operates at lower price levels but with visible fluctuations.
This is mostly reflected in the normalised IQR trajectory, where Arbitrum shows the most stable price, Base exhibits slightly higher variability, and Optimism variations are significantly more pronounced. 


The stability scores shown in the bottom right of Figure~\ref{fig:price_evolution} are derived from the IQR and the median gas price values.
Arbitrum is perhaps the most interesting case, as between~days 100 and 120, we observe a noticeable disruption in the~median price, which is also clearly reflected in the evolution of the stability score.

\observ{For the analysed time frame, Arbitrum is the most stable, Base lies in an intermediate range, and  Optimism shows the lowest stability.}

\section{Conclusions}
\label{sec:conclusions}

We addressed the problem of selecting an~appropriate~L2 rollup for deploying smart contracts by proposing a decision methodology.
Our approach takes into account multiple metrics, such as gas price, efficiency, popularity, and price~stability, and user profiles that reflect different deployment~priorities, enabling a flexible, multi-perspective evaluation of L2 solutions.

Using data collected from Arbitrum, Optimism, and Base and leveraging community-maintained smart contract categorisations, we computed scores for each contract category and each blockchain.
These scores highlight the suitability of the different rollups for deployment under varying assumptions.
Our results show that no single L2 dominates across all categories; instead, each rollup exhibits strengths in specific categories, emphasising the importance of context-aware decision-making.
In addition, our analysis of gas price evolution over a six-month period 
provides further insight into the predictability of transaction fees.

We note that our work and results are not intended to provide an absolute indication of deployment choice, but to offer
practical guidance for developers and stakeholders on the steps of choosing an appropriate L2 solution.
Results may vary consistently depending on the monitored period and the adjustment of profile weights.

As future work, we plan to extend our framework by adding additional metrics and incorporating more L2 solutions.
We aim to leverage this approach to design a more advanced engine that can better predict the most suitable deployment environment for smart contracts and, potentially, recommend migration paths between L2 solutions when appropriate.
Finally, the six-month observation window (September 2025 to March 2026) may reflect transient market conditions, and the resulting rankings could shift over longer periods. We plan to extend this window and compare the resulting rankings against those reported here.





\section*{Acknowledgments}
This work was supported by the \textit{CHIST-ERA-23-SMARTC-01-SCEAL} project, by a grant from the Romanian Ministry of Research, Innovation and Digitization, CNCS/CCCDI - UEFISCDI, project nr. 86/2025 ERANET-CHISTERA-IV-SCEAL, within PNCDI IV.


\bibliographystyle{IEEEtran}
\bibliography{refs}

\end{document}